\documentstyle[]{l-aa}

\begin{document}
\input{psfig.sty}

\thesaurus{10.07.2; 11.05.1}

\title{Halo and Bulge/Disk globular clusters in the S0 galaxy NGC 1380 
\thanks{Based on data collected at the European Southern Observatory,
         Chile}
}

\author {Markus Kissler-Patig \inst{1,2}, Tom Richtler \inst{1}, 
Jesper Storm \inst{3}, Massimo Della Valle \inst{4} 
} 

\offprints {mkissler@ucolick.org}

\institute{
Sternwarte der Universit\"at Bonn, Auf dem H\"ugel 71, 53121 Bonn, Germany
\and
UCO/Lick Observatory, University of California, Santa Cruz, CA 95060, USA
\and 
European Southern Observatory, Casilla 19001, Santiago 19, Chile
\and
Dipartimento di Astronomia, Universita di Padova, Vicolo Osservatorio 5, 
35122, Padova
}

\date {}

\maketitle

\begin{abstract}
We investigated the globular cluster system of the S0 galaxy NGC 1380 in
the Fornax cluster with deep $BVR$ photometry. We identified two,
presumably
old, populations of globular clusters. The blue globular clusters seem to
be counterparts to the halo globular clusters in our Milky Way. They have
comparable colors and magnitudes, are spherically distributed around
the elliptical galaxy but have a very flat surface density profile.  
The red population follows the stellar light in ellipticity and position
angle, has a similar magnitude distribution to the blue clusters, and a surface
density profile comparable to that of globular clusters in other 
Fornax ellipticals.
From their colors the red clusters appear slightly more metal rich then
the metal rich globular clusters in the Milky way. We associate this
red population with the bulge and disk of NGC 1380. While these two
populations are compatible with a merger formation, we unfortunately
see no hint in favor nor against a past merger event in the galaxy.
 
This would be the first identification of halo and bulge globular clusters
in an early--type galaxy by their colors and spatial distribution, and could 
hint to the presence of halo {\it and} bulge globular clusters in all galaxies.

The globular cluster luminosity functions in the three colors will be discussed 
in detail in a separate paper together with the implications of the absolute
magnitude of SN 1992A.


\keywords{globular cluster systems -- globular clusters -- early--type
galaxies }

\end{abstract}


\section{Introduction}
The astrophysical interest in investigating extragalactic 
globular cluster systems 
is mainly fed from two sources: first the relation
between properties of the globular cluster systems and properties of the host 
galaxies can be
interpreted in the context of galaxy evolution and may provide information
about merger history, bulge/halo populations, formation of globular
cluster systems, Dark Matter
content etc. For recent review articles and data compilations, see Harris \&
Harris (1996), Richtler (1995), and Ashman \& Zepf (1997).

Second, the globular cluster luminosity function can be calibrated
as an extragalactic distance indicator (e.g.~Jacoby et al.~1992, Whitmore 1996)
and enables distance determinations with HST out to a hundred Mpcs. 

NGC 1380 is an S0 galaxy in the Fornax cluster (see Tab.~1 for
general information). It was selected for observation as part of a study of the
applicability of the turn-over magnitude of the globular cluster luminosity
function as a distance
measure, in particular for studying the dependence of turn-over magnitude
on Hubble type. 
Furthermore, the Fornax cluster provides an important
stepping stone for the extragalactic distance scale, by hosting three well
observed supernovae of Type Ia: SN80D and SN81N in NGC 1316, and SN92A in 
NGC 1380.
SN92A ranks among the best spectroscopically and photometrically observed 
(Kirshner
et al.~1993, Suntzeff 1996). The distance scale aspects of this work 
will be detailed in a separate paper (Della Valle et al. 1997).

In the present paper we concentrate on the 
globular cluster system as a subpopulation of
its host galaxy. Due to the quality of the presented data, this objective
is very interesting particularly with regard to the S0 nature of the galaxy 
and to the comparison with other E and S0 galaxies in Fornax, 
which have been recently studied by Kissler-Patig et al.~(1997). 
One of the most interesting aspects of the data 
is the color distribution of globular clusters since only
a few globular cluster systems have been studied deeply in 3 colors, as 
is the case with the present data (Sect.~3). 
The total number of globular clusters and the specific frequency
are derived in Sect.~4.
A blue and a red population of globular clusters are identified, and
their properties are discussed in Sect.~5. 
Since S0 galaxies can be understood as being a transition form between
elliptical and spiral galaxies (e.g~Nieto \& Bender 1989), their globular cluster
systems might provide additional
clues concerning their position in the Hubble sequence. Our conclusions
in this respect are drawn in Sect.~6.   

\begin{table*}
\caption{\label{n1380} General data of our target galaxy, taken from de Vaucouleurs et
al.~(1991) and Poulain (1988)}
\begin{tabular}{l c c c c l l c c c }
\hline
name & RA(2000) & DEC(2000) & $l$ & $b$ & type & $m_V$ & $B-V$ & $V-R$ &
$V_{0}$[km~s$ ^{-1}$] \\
NGC 1380 & 03 36 27 & $-$34 58 33 & 235.93 & $-$54.06 & S0 & 11.81 & 0.98 & 0.57 &
1841 \\
\hline
\end{tabular}
\end{table*}
\begin{figure}
\psfig{figure=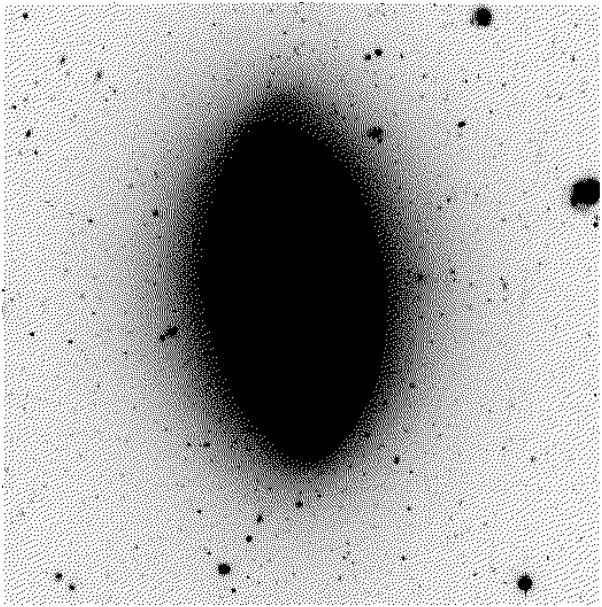,height=8cm,width=8cm
,bbllx=15mm,bblly=45mm,bburx=200mm,bbury=230mm}
\caption{
NGC 1380, $B$ image, 15000 seconds 
}
\end {figure}


\section{Observation and reduction}

The $B$, $V$, and $R$ images were obtained at the 3.5m New Technology
Telescope (NTT) at La Silla with the ESO Multi-Mode Instrument (EMMI).
The exposure times were chosen to reach approximately the same magnitude
beyond the expected turn-over in all three bands (see Tab.~2).
Thus the stacked frames have equivalent exposure times of 15000 seconds in $B$,
7800 seconds in $V$ and 2400 seconds in $R$ with effective seeing values
of $1.05''$, $1.00''$ and $0.95''$ respectively. 
A summary of the exposures is presented in Tab. \ref{tab.datalog}.
The sizes of the fields are
$4.98$ by 4.98 arcmin with a pixel size of $0.29''$ in $B$ using the blue
arm of EMMI,
and 6.29 by 6.29 arcmin with a pixel size of $0.37''$ in $V$ and $R$ where
the red arm of EMMI was employed.

\begin{table}
\caption{\label{tab.datalog} Summary of the NGC1380 observations
at the ESO NTT.}
\begin{tabular}{l c c c}
\hline\\
Date & Filter & Exposure & Number \\
  & & time (sec) & of exposures \\
\hline\\
04/09/92 & V & 600 & 1 \\
19/09/92 & V & 600 & 2 \\
15/11/92 & V & 600 & 1 \\
15/11/92 & R & 600 & 1 \\
16/11/92 & V & 600 & 3 \\
16/11/92 & R & 600 & 3 \\
22/11/92 & B & 1200 & 2 \\
22/11/92 & B & 1800 & 7 \\
22/11/92 & V & 600 & 6 \\
\hline\\
\end{tabular}
\end{table}

Isophotal models of the galaxy were built with the
ISOPHOTE package under IRAF, and subtracted from the original images in
order to get flat, homogeneous backgrounds for the point source search.
The photometry was done with DAOphot under IRAF. The DAOfind limit was set
to three standard deviations of the sky noise value. Between 2100 and 2600
objects were found in each color. Completeness calculations were done
with artificial star experiments, adding 200 artificial stars per run
over 100 runs in each color. Our 60\% completeness limits in individual
colors are $B=25.8$ mag, $V=24.7$ mag, $R=24.1$ mag. 

The calibration was done using eight local standards around NGC 1380
which were reasonably isolated and well exposed. Our calibration relation 
allowed us to recover the magnitudes of the local standards with an accuracy of 
0.021 mag in $B$, 0.016 mag in $V$ and 0.006 mag in $R$.
Our calibrated magnitudes were finally compared to the local standards used
in the SN campaign (Cappellaro et al.~1997)
and found to have a systematic offset of 
$0.07\pm0.01$ mag in $B$, $0.08\pm0.01$ mag in $V$, and $0.095\pm0.015$
 mag in $R$.  We decided to match our calibration with the one used for 
the SN, and applied the shifts to our measured magnitudes. Note that this
shift is negligible for the colors, number of globular clusters
and morphological properties, i.e.~the main issues of this paper. 

We combined our $B$, $V$ and $R$ object samples and obtained a list of 908
objects detected in the three colors. Clearly extended objects in this
sample were rejected using the chi and sharp values returned by DAOphot, to
leave us with 710 ``point sources''. We selected these in colors 
($0.7<B-R<1.9$, 468 bona fide globular cluster candidates, see also Sect.~3)
to derive morphological properties 
and the globular cluster luminosity function. And we selected the point 
sources with errors smaller than 0.1 mag in $B-R$ and 
$B-V$ (328 objects) to investigate the color distributions.


\section{Globular cluster colors}

In this section we discuss the colors of the point sources on our frame.
We assume $E(B-V)=0.0$ towards Fornax (Burstein \& Heiles 1982). Further,
Kirshner et al.~(1993) derived a reddening towards supernova 1992A in NGC 1380
of $E(B-V)\simeq0.0$, on the basis of the missing interstellar $UV$--lines 
in their spectra.
We are therefore confident that our colors are affected by internal
reddening in NGC 1380 only, if at all (see below).

We plotted the colors of all point sources with errors in $B-V$ and
$B-R$ smaller then 0.1 mag in Fig.~2. On the left hand side we plotted 
$B-V$ versus
$B-R$, and $V$ versus $B-R$ for NGC 1380, on the right hand side the
equivalent plots for the Milky Way globular clusters (de--reddened according
to the values given in Harris 1996, and for an arbitrary distance modulus 
of 31.0).
Figures 3 and 4 show the histograms over $B-V$
and $B-R$ for the same sample. We rejected a dozen from over 300 objects
that have colors
bluer than the bluest Milky Way globular clusters ($B-R<0.8$) as potential
contamination by foreground stars or background galaxies. Further, a clear gap
is visible at $1.9 < (B-R) < 2.2$ mag. Objects
redder than $B-R=2.0$ are most likely background galaxies, with about a
third of them associated with a background cluster (see
Sect.~3.2). Finally, what might appear to the eye as a potential parallel 
sequence in the color--color diagram of NGC 1380 is span by less than 
20 of the
300 objects, and might be a combination of photometric errors and
low-redshift background galaxies. No systematics in positions, colors or
magnitudes were found for this sample, that might hint to 5--10\% residual
contamination of our globular cluster sample. 
We focus in the following section onto point sources with colors
between $0.7 < (B-R) < 1.9$ mag, that we estimated to be good globular 
clusters candidates.
\begin{figure}
\psfig{figure=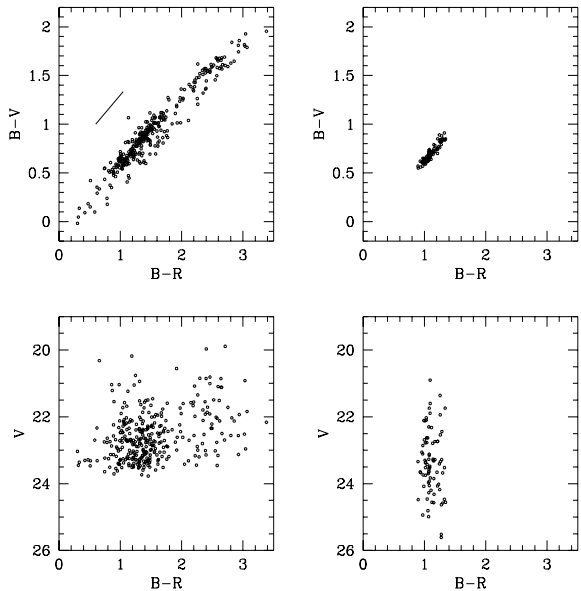,height=8cm,width=8cm
,bbllx=8mm,bblly=57mm,bburx=205mm,bbury=245mm}
\caption{
Color and magnitude comparison between NGC 1380 and Milky Way globular clusters.
On the left hand side we show a color--color and a color--magnitude
diagram for NGC 1380, on the right hand side the corresponding diagrams
for the Milky Way. The left upper panel includes a reddening vector
according to Cardelli et al.~(1989).
All point sources around NGC 1380 with $B-R$ and
$B-V$ errors less than 0.1 mag were plotted.
}
\end {figure}
\begin{figure}
\psfig{figure=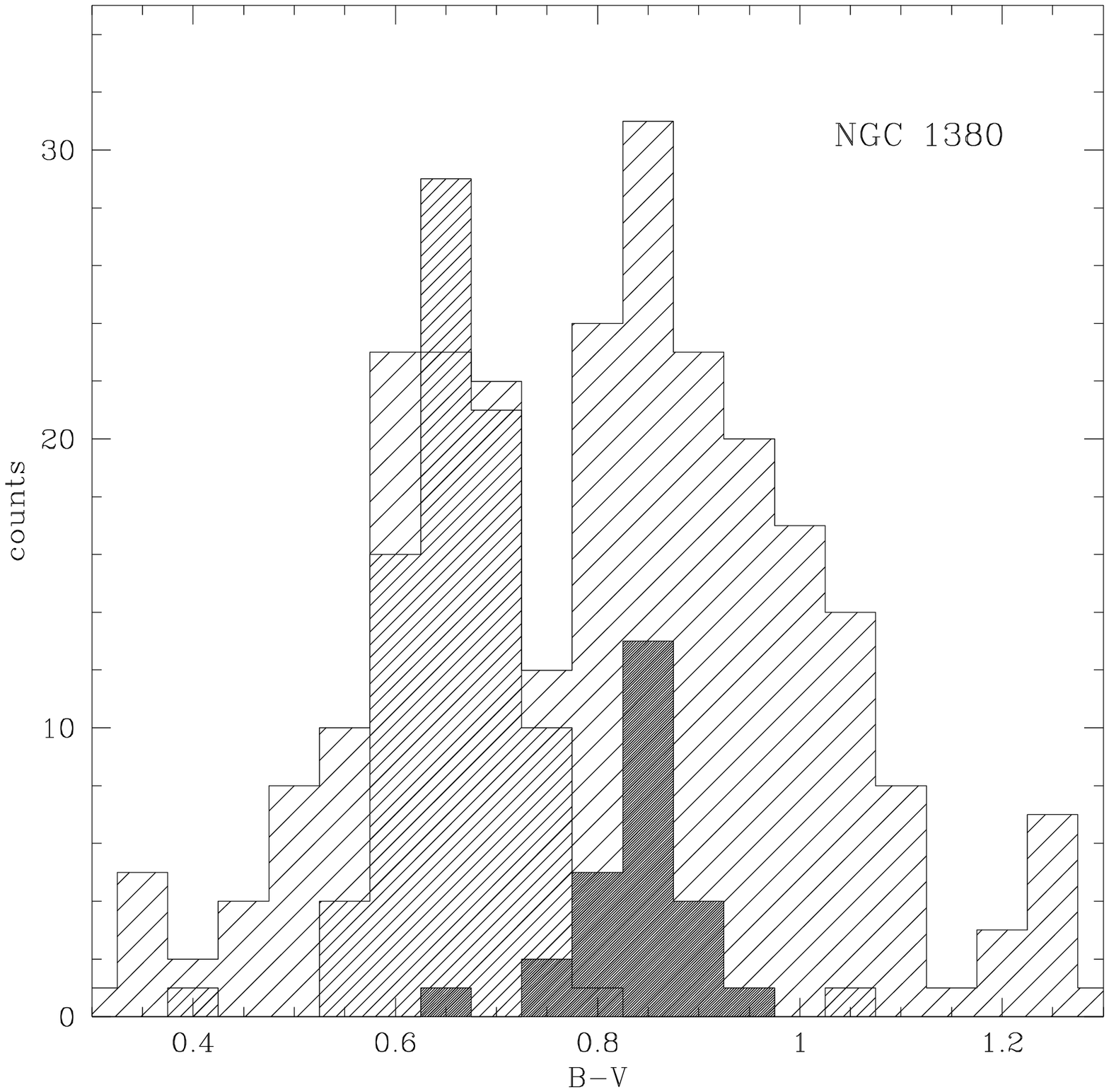,height=8cm,width=8cm
,bbllx=8mm,bblly=57mm,bburx=205mm,bbury=245mm}
\caption{$B-V$ color histogram of point sources around NGC 1380. Globular
clusters have colors between $0.4<B-V<1.2$, objects with $B-V>1.3$ are
most likely background galaxies. Overplotted are the colors
of metal--poor (grey shaded) and metal rich (dark shaded) Milky Way
globular clusters.
}
\end {figure}
\begin{figure}
\psfig{figure=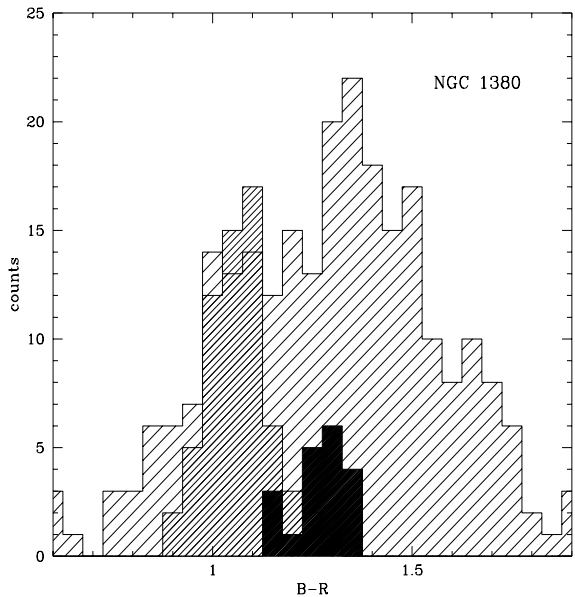,height=8cm,width=8cm
,bbllx=8mm,bblly=57mm,bburx=205mm,bbury=245mm}
\caption{
$B-R$ color histogram of point sources around NGC 1380. Globular
clusters have colors between $0.7<B-R<1.9$, objects with $B-R>2.0$ are
most likely background galaxies. Over-plotted are the colors
of metal--poor (grey shaded) and metal rich (dark shaded) Milky Way
globular clusters.
}
\end {figure}

\subsection{Halo and bulge globular clusters?}

Both color distributions ($B-V$ and $B-R$) are broader than
expected from our errors in the photometry alone ($< 0.1$ mag). A KMM
test (Ashman et al.~1994) rejects a unimodal distribution in favor of
two Gaussians with more than 96\% confidence for both colors.
The two peaks are identified at $(B-V)=0.65$ and 0.94 mag and $(B-R)=1.08$
and 1.52 mag. As a comparison, we divided the Milky Way globular
clusters into metal--poor ([Fe/H]$<-$0.9 dex) and metal--rich
([Fe/H]$>-$0.9 dex) objects, and over--plotted their color distributions (double
hashed and black histogram respectively) in Figs.~3 and 4.

The blue globular clusters in NGC 1380 match exactly the metal--poor
globular clusters of the Milky Way in both colors. Further, these blue
clusters are not brighter than expected (see below) and there is no evidence
that NGC 1380, an S0 galaxy, has undergone an interaction or a merger in
the last several Gyrs (see Sect.~6.1). We therefore reject the possibility 
that the blue color of the clusters is due to a younger age,
and are confident that they are the counterpart in NGC 1380 of 
the halo globular clusters in our Milky Way (see Sect.~6.2). 

The red globular clusters, on the other hand, are partly matched in
colors
by the metal--rich clusters of the Milky Way, but extend to slightly
redder colors. This could be due to different effects. First, 
we did not include in our plot Milky Way clusters that could not be
accurately dereddened. These mostly include metal--rich clusters close to
the plane of the galaxy (e.g.~Zinn 1985), i.e.~our comparison 
sample suffers some incompleteness at the red end.  But this can certainly 
not explain the whole red tail. Second, we could consider some
internal reddening in NGC 1380. However, the mean and median magnitudes 
differ by less then 0.2 mag in $V$ when we divide the red clusters into
two samples at $(B-R)\simeq1.45$ (see also Sect.~3.1, Fig.~6), that is, the 
reddest globular clusters are not
dimmed by extinction, making significant internal reddening unlikely
as already mentioned above. Finally, assuming these clusters to be roughly of
the same age as the Milky Way clusters, the red color would be a metallicity
effect. If
we extrapolate a color--metallicity relation for the Milky Way
([Fe/H]=$-5.4(\pm0.3) + 3.7(\pm0.3)(B-R)$, using the compilation of
Harris 1996), the mean color of the red globular clusters
$B-R\simeq 1.50$ would lead to a metallicity of about solar, as observed in 
elliptical galaxies (e.g.~ the compilation of Ashman \& Zepf 1997).

The magnitude distributions of the red and blue clusters are similar. 
We show in Fig.~6 the
$B$ globular cluster luminosity functions for the blue, the red, and all
 globular clusters located at distances from 35'' to 150'' from the
center of NGC 1380.
\begin{figure}
\psfig{figure=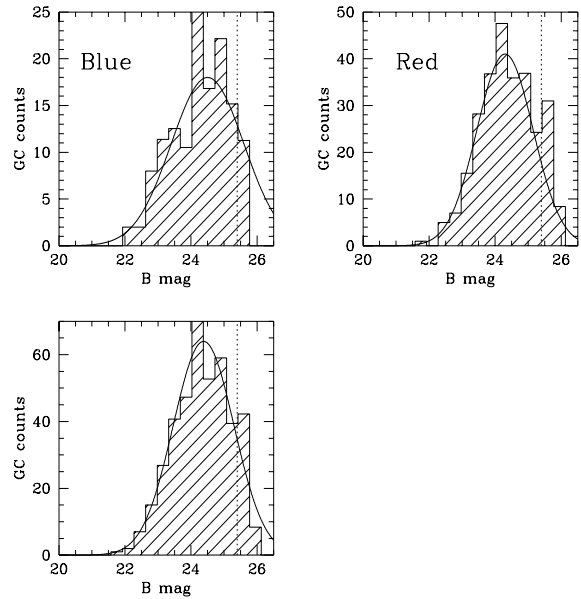,height=8cm,width=8cm
,bbllx=8mm,bblly=57mm,bburx=205mm,bbury=245mm}
\caption{
Globular cluster luminosity functions in $B$ for blue and red globular clusters
around NGC 1380 (upper panels), and for the combined sample (lower
panel). The best Gaussian fits are over--plotted, and the dotted line
marks the 50\% completeness limit. 
}
\end {figure}
The best Gaussian fits to the data return turn--over luminosities of
$m_B^{TO}=24.5\pm0.15$ mag and $m_B^{TO}=24.3\pm0.15$ mag for the blue and red
samples respectively, and $m_B^{TO}=24.38\pm0.1$ mag for the combined
sample (see also Della Valle et al.~1997). In $V$ and $R$ the
turn--overs still agree with the changing mean colors of the sample
(i.e.~ $m_B^{TO}-m_V^{TO}=0.7\pm0.2$, $m_B^{TO}-m_R^{TO}=1.1\pm0.2$ for the
blue clusters; $m_B^{TO}-m_V^{TO}=0.9\pm0.2$,
$m_B^{TO}-m_R^{TO}=1.3\pm0.2$ for the red clusters).
These results agree well with the values found for other small
early--type galaxies in Fornax (Kohle et al.~1996), indicating from
their magnitudes that our two populations are likely to be old.  Note however, that
metal--poor (blue) globular clusters should be brighter than metal--rich
(red) globular cluster at the same age (e.g.~Ashman et al.~1995), and the
effect should increase toward bluer colors. In contrast, our data hint to a
systematicly brighter {\it red} sample, even if the effect is hard to 
disentangle from our errors. Assuming mean metallicities similar to the halo and
the bulge of the Milky Way, magnitude differences between blue and red
clusters should be of the order of $\Delta B\simeq 0.35$, $\Delta V\simeq
0.25$, and $\Delta R\simeq 0.20$, according to Ashman et al.~(1995).
This is just outside the range compatible with our errors, and we conclude
that some age effect is compensating for the metallicity effect and
be brightening our red clusters, i.e.~the red globular clusters may be
slightly younger than the blue ones. According to the population
synthesis models of
Fritze-v.Alvensleben \& Burkert (1993), the red cluster would have to be
about 3 Gyrs younger on average than the blue ones. Worthey's
(1994) models return a similar age difference
of less than 4 Gyrs (assuming in both cases a similar globular cluster
mass distribution for the red and blue clusters).

In summary, NGC 1380 appears on the basis of colors to have two  distinct
populations of old globular clusters, which will be confirmed by their
spatial distribution in Sect.~5. The blue globular clusters are very similar
to the metal--poor counter--parts in the Milky Way halo. The red globular 
clusters are, in the mean, slightly redder, i.e.~most likely more metal--rich
than the metal--rich population in the Milky Way. Further, there is weak
evidence that the red population could be 3-4 Gyrs younger than the blue one.

\subsection{A background galaxy cluster}

Figure 6 shows the location of objects with colors between $2.1 < (B-R)
< 2.7$ mag. More than a third of them are concentrated in a group 30"
east of the center of NGC 1380. These objects have $V$ magnitudes
between 21.0 and 22.9 mag, and so cannot be reddened globular
clusters, which would appear more than 1 magnitude fainter than the brightest
blue objects ($V>22.0$) if suffering from extinction. 
This supports the conclusion already mentioned above that there is little 
internal extinction in NGC 1380. Further, these objects fit the K--corrected colors and magnitudes
of early--type galaxies at $z=0.5\pm0.1$ (e.g.~Coleman et al.~1980). 
Finally we measured the Gaussian FWHM of these objects with  $22<B<23$ 
mag to be $1.14\arcsec \pm 0.02\arcsec$, compared to globular
cluster candidates in the same magnitude range that have $1.03\arcsec
\pm 0.02\arcsec$.  
We suggest that these objects are likely to belong to a galaxy cluster at 
intermediate redshift.
\begin{figure}
\psfig{figure=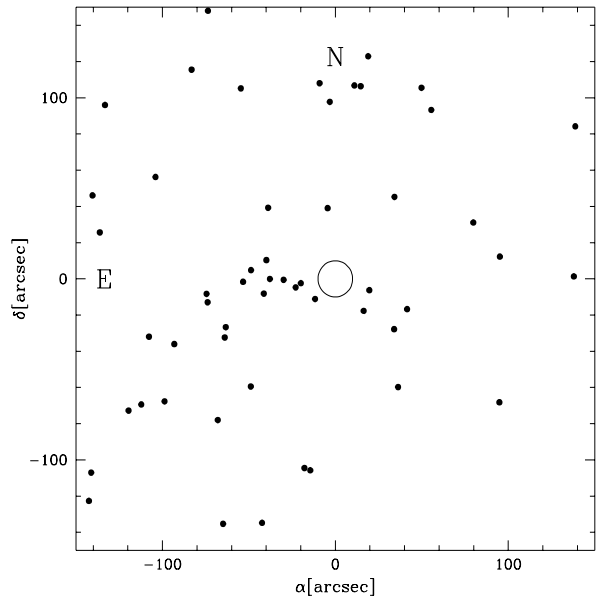,height=8cm,width=8cm
,bbllx=8mm,bblly=57mm,bburx=205mm,bbury=245mm}
\caption{
Distribution of the objects with $2.1<(B-R)<2.7$ around NGC 1380
}
\end {figure}

As a last consistency check  we consider all point sources with $B-R<2.0$ to be mostly globular
clusters, but all point sources with $B-R>2.0$, and {\it all} extended
sources to be
background galaxies, we end up with roughly 33500 background galaxies per
square degree down to $V=23.5$ (i.e.~231 objects on 24.8 square arcmin),
which compares well with e.g.~Smail et al.'s (1995) deep counts, that predict
about 33000 objects per square degree.
If, however, most point sources with $B-R>2.0$ were reddened globular
cluster, we would end up with less than 2/3 of the expected background
galaxies. 


\section{The number of globular clusters and specific frequency}

\subsection{The number of globular clusters}

The globular cluster system is fully covered by our frame, as shown by our
density profiles in Sect.~5.2. We therefore do not have to apply any
corrections for geometrical incompleteness and can concentrate on the
number of globular clusters found on our images.
 
One way to compute the number of globular clusters around NGC 1380 is
to integrate analytically our globular cluster luminosity functions 
(see Della Valle et al.~1997). We then get $555\pm33$ globular clusters,
averaged over the $B$, $V$, and $R$ globular cluster luminosity
functions.

Another method is to count the number of globular clusters (i.e.~point
sources with $0.7<(B-R)<1.9$ and $0.4<(B-V)<1.2$), down to our
turn--over magnitudes and multiply them by two, assuming the luminosity
function to be symmetric around the turn--over magnitude. We get
$564\pm20$ objects as a mean from counts down to 
$m^{TO}_B=24.38$ mag, $m^{TO}_V=23.67$ mag, and $m_R ^{TO}=23.16$ mag,
multiplied by two.

If we divide our sample by color as proposed in the last section, we get
$191\pm10$ red globular clusters and $91\pm10$ blue globular clusters
from counts down to the same turn--over magnitudes. This is a total 
of $382\pm20$
red and $182\pm20$ blue globular clusters around NGC 1380, and a ratio
of roughly 2 to 1 red to blue globular clusters.

\subsection{The specific frequency}

In order to derive the specific frequency, we need to adopt a distance
modulus for NGC 1380. Here we use the distance modulus derived from the 
globular cluster luminosity function (see Della Valle et al.~1997) of
$(m-M)=31.35\pm0.16$, in excellent agreement with the distance modulus 
derived from Cepheids by the HST Key project for NGC 1365 
(another Fornax member) of $(m-M)=31.32\pm0.19$.
The visual $V$ magnitude of NGC 1380 is $m_V=9.91\pm0.10$ mag
(see Sect.~1). The specific frequency, using
$M_V=-21.44\pm0.21$ and $N_{GC}=560\pm30$, is then $S_N=1.5\pm0.5$. 
As a comparison, the mean for elliptical galaxies in Fornax (excluding NGC 1399)
is $S_N=4.0\pm1.0$
(Kissler-Patig et al.~1997), the neighbour S0 galaxy NGC 1387 has an 
$S_N=3.2\pm1.1$ (Kissler-Patig et al.~1997), and $S_N$ is known to
decrease with later types (Harris 1991).  NGC 1380, an S0 type 
galaxy, has a somewhat low but still normal specific frequency for a S0
galaxy in a clusters.


\section{Morphological properties}

In the following section we investigated the ellipticity and position
angle, as well as the surface density profile of the globular cluster system.
These properties are then compared with those of the host galaxy light. 
We used the sample defined in Sect.~2, i.e.~all point sources found on all
three frames, selected in color. Further, to avoid any errors introduced by
completeness corrections,  we considered only
objects brighter than the 80\% completeness limit in all three colors.

\subsection{Ellipticity and position angle}

To determine the ellipticity and position angle of the globular cluster
system, we further considered only globular clusters far enough from the center ($\simeq
50\arcsec$) not to suffer from increased detection incompleteness, and close 
enough ($< 150\arcsec$) to lie within a ring totally enclosed on the
frame in order to avoid geometrical completeness corrections. We were
left with a sample of 220 bona fide globular clusters.

These objects were then divided into sectors 30\degr wide around the
center of NGC 1380. The result is plotted in Fig.~7. 
In case of a spherical distribution, we would expect a 
constant number of globular clusters at all position angles. Instead we
see two peaks at the position angle of the galaxy, representing an 
elongated globular cluster system.
The best fit of a double cosine returns a position angle of $3\degr \pm10\degr$ 
and an ellipticity of $\epsilon_{\rm GCS}=0.5\pm0.2$ (following our method
applied in Kissler-Patig et al.~1996), compared to  
a position angle of $5\degr\pm5\degr$ and an ellipticity of
$\epsilon=0.45\pm0.10$ for the galaxy, taken from our isophotal models
in $B, V,$ and $R$. Note that we get very similar results for the
properties of the globular cluster system when changing the sector width,  
or using the objects from individual colors only.
\begin{figure}
\psfig{figure=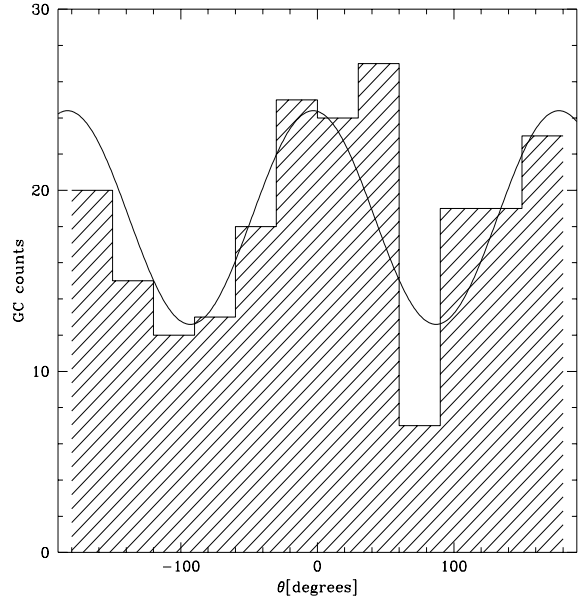,height=8cm,width=8cm
,bbllx=8mm,bblly=57mm,bburx=205mm,bbury=245mm}
\caption{The angular distribution of the globular clusters around NGC
1380 in 30\degr sectors. The over--plotted solid line is the best
fit to the data, and coincides in position angle and ellipticity with
the galaxy light.
}
\end {figure}
The globular cluster system of NGC 1380 seems to follow
the ellipticity and position angle of the galaxy light.

However, if we divide our objects into blue and red clusters as
discussed in Sect.~3, and plot their angular distribution (see Fig.~8,
we applied a point symmetry around the center of NGC 1380 to increase the
counts in each bin), we note that the red sample follows the galaxy in
ellipticity and position angle, but that the blue globular clusters are
rather
spherically distributed around the galaxy (constant number of clusters at all
angles). While the red clusters seem to be associated with the galaxy
light, the blue clusters appear to be morphologically decoupled, which
indicates a distinction in halo and bulge/disk globular clusters. 
\begin{figure}
\psfig{figure=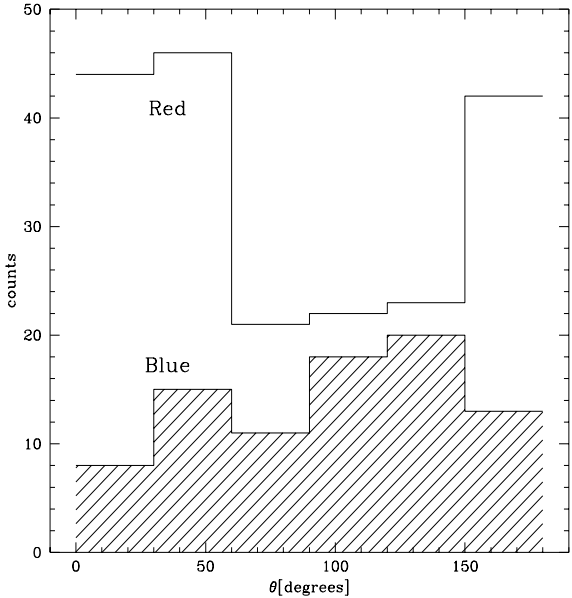,height=8cm,width=8cm
,bbllx=8mm,bblly=57mm,bburx=205mm,bbury=245mm}
\caption{The angular distribution of halo and bulge globular clusters around NGC
1380 in 30\degr sectors, after a point symmetry around the center of
the galaxy. 
}
\end{figure}

\subsection{The surface density profile}

The surface density profile of the globular cluster system was 
investigated on the $B$,$V$, and $R$ images individually in order to be 
compared with the galaxy light profile in the respective color.
We used all objects (point sources and extended objects) found on the frame
down to the 80\% completeness limit in the individual colors
($B<25.5$, $V<24.4$, $R<23.9$).
We computed the number of globular clusters in 16 elliptical ($\epsilon_{\rm GCS}
= 0.5$ see last section) rings, 14.5\arcsec\ wide on the $B$ image, 
18.5\arcsec \ wide on the $V$, and $R$ image. The last five rings suffer
from more than 20\% geometrical incompleteness, the last ring is only
covered to 30\% by our frame, and thus should be regarded with caution.
The mean major axis and corresponding surface density with its error are shown 
in Tab.~3. The profiles are plotted in Fig.~9. The right
panels show the raw counts with our estimated background as dotted line.
The left panels show the corrected surface density of the globular clusters,
together with the light profile of the galaxy in the respective filter,
taken from our isophotal models. 

\begin{table}
\caption{Surface density profiles in $B$, $V$, and $R$. The individual
columns for each color list the mean major axis in arcsec and the
corresponding surface density with its error in counts per square
arcmin}
\begin{tabular}{rc rc rc}
\hline
\multicolumn{2}{c}{$B$ data} & \multicolumn{2}{c}{$V$ data
} & \multicolumn{2}{c}{$R$ data} \\
$a$ & density & $a$ & density & $a$ & density \\
\hline
  7  & $456.9 \pm  67.3$&    9 & $ 334.8 \pm 45.1$ &   9 & $ 140.0\pm   29.1$\\
 22  & $313.5 \pm  32.1$&   28 & $ 229.2 \pm 21.5$ &  28 & $ 190.7\pm   19.6$\\
 36  & $212.0 \pm  20.5$&   46 & $ 122.9 \pm 12.2$ &  46 & $ 131.4\pm   12.6$\\
 51  & $125.9 \pm  13.3$&   65 & $  91.3 \pm 8.9 $ &  65 & $ 108.7\pm    9.7$\\
 65  & $116.7 \pm  11.3$&   83 & $  62.2 \pm 6.4 $ &  83 & $ 71.7 \pm   6.9$\\
 80  & $85.5  \pm   8.7$&  102 & $  49.2 \pm 5.2 $ & 102 & $ 48.1 \pm  5.1$\\
 94  & $73.1  \pm   7.4$&  120 & $  40.7 \pm 4.3 $ & 120 & $ 37.4 \pm  4.1$\\
109  & $76.6  \pm   7.1$&  139 & $  28.8 \pm 3.4 $ & 139 & $ 30.0 \pm  3.4$\\
123  & $51.2  \pm   5.4$&  157 & $  29.7 \pm 3.2 $ & 157 & $ 27.5 \pm  3.1$\\
138  & $54.9  \pm   5.4$&  176 & $  27.5 \pm 3.0 $ & 176 & $ 25.8 \pm  2.9$\\
152  & $54.2  \pm   5.6$&  194 & $  26.4 \pm 3.2 $ & 194 & $ 23.4 \pm  2.8$\\
167  & $49.0  \pm   5.6$&  213 & $  28.5 \pm 3.4 $ & 213 & $ 26.2 \pm    3.2$\\
181  & $43.5  \pm   5.4$&  231 & $  20.8 \pm 3.0 $ & 231 & $ 24.2 \pm  3.1$\\
196  & $64.8  \pm   6.6$&  250 & $  21.2 \pm 3.1 $ & 250 & $ 22.4 \pm  3.0$\\
210  & $58.9  \pm   6.8$&  268 & $  30.8 \pm  3.7$ & 268 & $ 28.9 \pm  3.5$\\
225  & $61.5  \pm   7.8$&  287 & $  20.0 \pm 3.0 $ & 287 & $ 20.2 \pm  2.9$\\
\hline
\end{tabular}
\end{table}
We fitted the globular cluster surface density profiles with a power law
of the form $\rho \sim r^{x}$, where $\rho$ stands
for the surface density of the globular clusters or the surface
intensity of the galaxy light, and $r$ for the semi--major axis
of both the globular cluster density
profile and the light of the galaxy (taken from our isophotal models,
see Sect.~2). The profiles were fitted over different radii (starting between
0\arcsec\ and 30\arcsec and extending out between 80\arcsec\ and
160\arcsec), while varying the background estimate by up to 50\% around the
mean of the last 8 bins of Tab.~3. We get slopes of $x=-1.55\pm0.5,
-1.63\pm0.08,$ and $-1.60\pm0.20$ in $B,V,R$ respectively for the globular 
clusters where the errors reflect the range of values obtained by varying the
background level and the fitting range. For the galaxy light we obtained 
$x=-1.33\pm0.06$, and $-1.34\pm0.06$ in $B$
and $V$. Because of a lower signal--to--noise in $R$ we forced an isophotal
model with fixed angle on the image, distorting the profile and
making a fit unreliable in this filter.
The total globular cluster system appears to compatible with the density
profile of the galaxy light.

The galaxy profile appears clearly flatter than an $R^{1/4}$ profile, probably 
due to a significant disk to bulge ratio (see also Bothun \& Gregg
1990). We could not disentangle both components in our profile fits.
Note further that Horellou et al.~(1995) suggest from the absence of HI and 
weak CO emission, that NGC 1380 might have suffered tidal stripping.
If this happened recently enough, the galaxy  might still be kinematically
hot. 
\begin{figure}
\psfig{figure=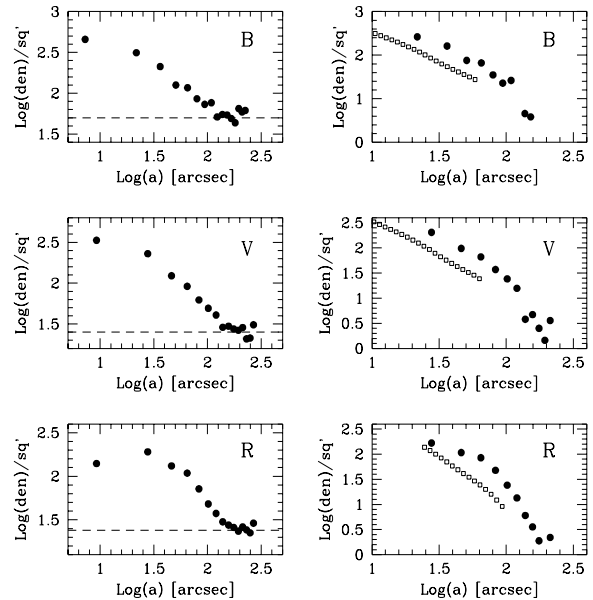,height=8cm,width=8cm
,bbllx=8mm,bblly=57mm,bburx=205mm,bbury=245mm}
\caption{Surface density profiles of all objects (left) and globular
clusters (right) around NGC 1380.
}
\end {figure}

We also investigated the surface density profiles of the red and blue 
population individually.  Their globular cluster
surface density profiles are plotted against radius and semi-major 
axis in Fig.~10.    
\begin{figure}
\psfig{figure=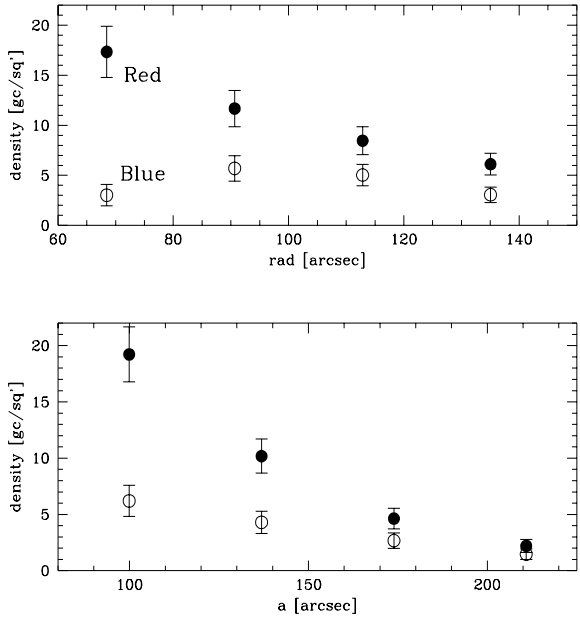,height=8cm,width=8cm
,bbllx=8mm,bblly=57mm,bburx=205mm,bbury=245mm}
\caption{Surface density profiles of red and blue globular clusters around 
NGC 1380, plotted once against the radius in arcseconds (upper panel)
and once against the semi-major axis (lower panel).
}
\end {figure}
We fitted the surface density profiles as described above (varying fitting
radii and background level) and obtained a
slope of $x=-2.5\pm0.3$ for the red objects, and $x=-1.0\pm0.4$ for the
blue objects. The blue objects follow a significantly flatter density profile 
than the red objects and even appear slightly flatter than the galaxy light. 
The red objects follow a steeper
profile than the galaxy light, but with a slope similar to that of the profiles
of globular cluster systems in small early--type galaxies 
(e.g.~Kissler-Patig 1997). The profile of the red objects agrees roughly
with a $R^{1/4}$ profile, and could be associated with the main stellar
component of NGC 1380.
The superposition of both red and blue profiles make the overall
globular cluster system look almost as flat as the galaxy light, as shown
above.


\section{The origin of the two populations}

We identified two populations of globular clusters in the S0 galaxy NGC 1380.
These two populations differ clearly in their spatial distribution and in
color but apparently not in their magnitude distribution. As explained in
Sect.~3.1, they must differ in metallicity and possibly somewhat in age.
Thus these two populations must have formed in different processes,
probably at slightly different epochs.

Two main alternatives for the presence of red and blue globular
clusters exist.\\
$\bullet$ Multi--modal color distribution in early--type galaxies 
could be associated with a merger event that induced the
formation of new, more metal--rich globular clusters (red population),
superimposed on the old, metal--poor (blue) populations of the
progenitors (e.g.~Ashman \& Zepf 1992, Zepf \& Ashman 1993).\\
$\bullet$ The bi--modal color distribution in the Milky Way corresponds
to the metal--poor globular clusters associated with the halo and
accretion of small companions (e.g.~Zinn 1996), and to the metal--rich
population associated with the bulge and disk.

\subsection{Did NGC 1380 form by merging?}

S0 galaxies are thought to be the continuation of the Hubble sequence
from disky ellipticals towards higher disk--to--bulge ratios
(e.g.~Kormendy \& Bender 1996). If low luminosity ellipticals formed by
mergers, S0's might be expected to have formed via merging too.
As a rather luminous early--type galaxy in a cluster, NGC 1380 is
expected to have formed at early times (e.g.~Bender, Burstein \& Faber 1993).
NGC 1380 appears as an isotropic oblate rotator (D'Onofrio et al.~1995) and
shows no substructure, but this does not exclude a merger.
NGC 1380 is not reported to show any signs for a recent merger event in its
history, and we find no anomalies (ripples or shells) except maybe for a
slightly warped disk, by subtracting our isophotal models.
The absence of detectable HI and CO (Horellou et al.~1995) is
interpreted as a hint to stripping but does not contradict an early
merger event either. We note however that NGC 1380 is the second brightest
early--type galaxy (after the central giant elliptical NGC 1399) in the 
Fornax cluster, and would most likely to be the product of a merger in
hierachical formation scenarios. However we are left with neither any clear
observational signs in favor nor against a past merger event for the
galaxy.

The globular cluster system, from the properties of blue and red globular
clusters, seem to be in agreement with the model of Ashman \& Zepf (1992),
with respect to the morphological properties (newly formed, i.e.~red
cluster, would be more centrally concentrated, the old clusters more
extended). The only open issue would be the low specific frequency
despite the high number of red (new) relative to blue (old) clusters. 
The low specific frequency indicates that no excess of globular
clusters (when compared to stars) was formed in the merger event, leaving
the specific frequency at a normal value after the merger.
However, on the other hand, the red population in NGC 1380 is more than twice 
as numerous as the blue one, implying that the merger event formed twice as
many globular clusters as were brought in by the progenitor galaxies. 
Both points combined imply that the merger was highly effective in forming
new stars. Unfortunatly no quantification is possible at this stage, but as
a comparison, the Antennae (NGC 4038/4039) currently forms 700 new globular 
clusters candidates from which many are expected to survive (Whitmore \& 
Schweizer 1995). The end product of these two Sb/Sc spirals is expect to
have a specific frequency around 2, and might be close 
to what NGC 1380 looks like today.

Therefore, we cannot rule out nor conclude that NGC 1380 is a merger
product, but if it formed in a merger, we note two interesting facts:
first, since the age of the red globular
clusters seem to be less than 4 Gyrs younger than the blue ones
(see Sect.~3.3.1), the merger must have happened at very early times.
Second, if all red globular clusters formed in the merger, the globular
clusters to star formation rate implies that mergers might not necessarely
change the specific frequency of galaxies, despite producing a high number
of new clusters.

\subsection{An analogy to the Milky Way?}

Alternatively, if NGC 1380 never experienced a major merger event, the two 
populations of globular clusters might be thought of the analogy of the halo and
bulge/disk globular cluster population in the Milky Way. 

Following characteristics are compatible with such a picture.
The magnitudes of blue and red globular clusters are similar
within few tenth of a magnitude, reducing the age difference to
a few Gyrs (see Sect.~3.3.1), i.e.~the spread also observed in the
Milky Way (Chaboyer et al.~1996).
From their luminosities and colors the two populations of globular
clusters in NGC 1380 appear old ($>12$ Gyrs, according to the model of
Worthey 1994). The blue globular clusters have luminosities and
colors comparable to the halo clusters in the Milky Way, and are spherically
distributed. NGC 1380 has a higher number of blue clusters ($180\pm20$) 
than the Milky Way ($\simeq 100$), but is also more luminous.
Further the red clusters follow closely the shape (ellipticity and position
angle) of the stellar component (mainly bulge), and also share the same
mean color with the stellar population ($B-V=0.94$ for the red globular
clusters (see Sect.~3.3.1),
compared to $B-V\simeq 0.98$ for the galaxy (see Tab.~1)).

However, such an analogy has its limits. The blue clusters have a far more 
flatter profile than the halo clusters in
the Milky Way. In the Galaxy, the halo population falls off spatially with
a power of -3.5, i.e.~in projection with a power of -2.5, distinctly
steeper than in NGC 1380. One could speculate that if indeed the galaxy 
suffered stripping lately (see above), this might account for the flatter 
profile of the outer globular clusters. However, we further do not 
expect the red globular
clusters to the younger in this analogy. There is now main sequence 
photometry for the most metal-rich galactic bulge clusters
(Ortolani et al.~1995, Fullton et al.~1995),
which indicates for these clusters an age comparable with that of halo 
clusters. The youngest galactic globular clusters
are not found among the bulge but among the halo clusters: Several objects
are known to be older by a few Gyrs than the bulk of halo clusters and it
is attractive to conjecture that they are the debris of accreted dwarf galaxies
(e.g.~Zinn 1996). This scenario in its simple form is not very
probable for NGC 1380, because dwarf galaxies should be metal poor and 
accordingly also their debris, and therefore enrich the blue population
with younger globular clusters.

Whatever the formation processes are, keeping this line of argumentation
implies for NGC 1380 a bulge formation {\it after} halo formation, which 
fits to some theories of bulge formation (Wise et al.~1997).

We note in passing, that it seems difficult to associate all red globular 
clusters in NGC 1380 with its disks as proposed in the Milky Way by Zinn (1996),
given their large number and their spatial distribution.
Alternatively at least part of the red globular cluster
population might be associated with the bulge 
as proposed for the Milky Way by Minniti (1995) and
Burkert \& Smith (1997).	

Finally we note that the two alternatives given above are not mutually
exclusive. On the one hand, an unanswered question about the Milky Way is
whether bulge and thick disk formed as a result of an interaction or
merger, making both hypothesis valid at the same time. On the other hand,
one could think of two red populations, namely one brought in by the
bulges/thick disks of the progenitors, and the other formed during the
merger event.

\section{Summary and conclusions}

We detected two clearly distinct populations of globular clusters in 
NGC 1380, pointing to two different globular cluster formation events in
this galaxy. 
To our knowledge, the identification of a population of halo clusters would
be the first identification of an halo component in a S0 galaxy. Of great help
is of course the unusually deep limiting magnitude in our study. Therefore the
comparison with GCSs of other elongated early type galaxies might remain vague.
Recently, Elson (1997) reported two distinct stellar populations in the S0 NGC
3115, however could not study their properties due to too small spatial
coverage.
Since S0 galaxies can be seen as the extension of early--types in the Hubble
sequence (Nieto \& Bender 1989), and given the similarity of the properties
of bulges of spirals and ellipticals (Bender et al.~1992, Wise et al.~1997),
blue ``halo'' globulars could be present in all early--type galaxies.
If elliptical galaxies are mostly bulge dominated,
the color distributions of the globular clusters in elliptical galaxies
will look unimodal and ``red'' as long as the halo population represents only a
small fraction of the total system. This could be the case in NGC 720
(Kissler-Patig et al.~1996), where no spherical component in the globular
clusters system could be detected.
The current detection limit of two populations largely
depends on the number ratio of the two populations, the accuracy of the photometry,
and the photometric system used. While multiple populations are detected in
several bright elliptical galaxies (e.g.~summary by Ashman \& Zepf 1997), 
the current broad band photometry
investigations can barely exclude ``unimodal'' color distributions in
normal, small elliptical galaxies, with two coeval populations differing only 
in metallicity (e.g.~Kissler-Patig et al.~1997), leaving the issue open.

The different formation processes of the two globular cluster populations
in NGC 1380 are not well enough constrained by our data to differenciate
between a merger formation or a Milky Way analogy. However we note that in
case of a merger formation, the merger event must have happened at very
early times, and despite producing a large number of new globular clusters,
it did not increase the specific frequency of the galaxy.
Finally, the blue globular clusters, despite many similarities to the Milky
Way halo clusters, have a much flatter radial profile. 


\acknowledgements

We thank D.~Minniti, D.~Forbes , S.~Zepf and S.~Ortolani for fruitful 
conversations and helpful comments.  MKP acknowledges the DFG Graduierten
Kolleg ``Das Magellansche System und andere Zwerggalaxien'' for a stipend. 
TR acknowledges the support of the DFG project Ri 418/5-1.


\begin{thebibliography}{}

\bibitem[]{}
Ashman K., Zepf S.E., 1992, ApJ 384, 50
\bibitem[]{}
Ashman K.M., Bird C.M., Zepf S.E., 1994, AJ 108, 2348
\bibitem[]{}
Ashman K.M., Conti A., \& Zepf S.E., 1995, AJ 110, 1164
\bibitem[]{}
Ashman K., Zepf S.E., 1997, ``Globular Cluster Systems'', Cambridge
University Press
\bibitem[]{}
Bender R., Burstein D., Faber S.M., 1992, ApJ 399, 462
\bibitem[]{}
Bender R., Burstein D., Faber S.M., 1993, ApJ 411, 153
\bibitem[]{}
Blakeslee J.P., 1996, PhD thesis, Massachusetts Insitute of Technology
\bibitem[]{}
Bothun G.D., Gregg M.D., 1990, ApJ 350, 73
\bibitem[]{}
Burkert A., Smith G.H., ApJ Letters in press
\bibitem[]{}
Burstein D., Heiles C., 1982, AJ 87, 1165
\bibitem[]{}
Cappellaro et al., 1997, preprint
\bibitem[]{}
Cardelli J.A., Clayton G.C., Mathis J.S., 1989, ApJ 345, 245
\bibitem[]{}
Chaboyer B., Demarque P., Sarajedini A., 1996, ApJ 459, 558
\bibitem[]{}
Coleman G.D., Chi-Chao Wu, Weedman D.W., 1980, ApJS 43, 393
\bibitem[]{}
Della Valle et al., 1997, in preparation
\bibitem[]{}
D'Onofiro M., Zaggia S.R., Longo G., et al., 1995, A\&A 296, 319
\bibitem[]{}
Elson R., 1997, MNRAS in press
\bibitem[]{}
Fritze-v.~Alvensleben, U., Burkert, A., 1993, A\&A 300, 58
\bibitem[]{}
Fullton L.K., Carney B.W., Olzewski E.W., et al.~1995, AJ 110, 652
\bibitem[]{}
Geisler, D., Piatti, A. E., Claria, J. J., \& Minniti, D. 1995, AJ, 109,
605
\bibitem[]{}
Geisler D., Lee M.G., Kim E., 1996, AJ 111, 1529
\bibitem[]{}
Harris W.E., 1991, ARA\&A 29, 543
\bibitem[]{}
Harris W.E., 1996, AJ 112, 1487
\bibitem[]{}
Harris W.E., Van den Bergh S., 1981, AJ 86, 1627 
\bibitem[]{}
Harris H.C.,\& Harris W.E.,  1996, Astrophysical Quantities, 4th edition
\bibitem[]{}
Horellou C., Casoli F., Dupraz C., 1995, A\&A 303, 361
\bibitem[]{}
Kirshner R.P., Jeffery D.J., Leibundgut B., et al., 1993, ApJ 415, 589
\bibitem[]{}
Kissler-Patig M., 1997, A\&A 319, 83
\bibitem[]{}
Kissler-Patig M., Richtler T., Hilker M., 1996, A\&A 308, 704
\bibitem[]{}
Kissler-Patig M., Kohle S., Richtler T., et al., 1997, A\&A 319, 470
\bibitem[]{}
Kohle S., Kissler--Patig M., Hilker M., et al.~, 1995, A\&A 309, L37
\bibitem[]{}
Kormendy J., Bender R., 1996, ApJ 464, L119
\bibitem[]{}
Minniti D., 1995, AJ 109, 1663
\bibitem[]{}
Nieto J.--L., Bender R., 1989 A\&A 215, 266
\bibitem[]{}
Ortolani S., Bica E., Barbuy B., 1995, A\&A 296, 680
\bibitem[]{}
Poulain P., 1988, A\&AS 72, 215
\bibitem[]{}
Richtler T., 1995, in ``Reviews of Modern Astronomy'', Vol.8,
eds.~G.~Klare, Springer, p.163
\bibitem[]{}
Smail I., Hogg D.W., Yan L., Cohen J.G., 1995, ApJ 449, L105
\bibitem[]{}
Suntzeff N.B., 1996, in Supernovae and Supernovae Remanents, IAU Colloq.~145, 
eds. McCray and Wang, Cambridge University Press, 41
\bibitem[]{}
Vaucouleurs de G., Vaucouleurs de A., Corwin H.G., Buta R.J.,
Paturel G., Fouqu\'e P., 1991, Third Ref.~Catalogue of Bright
Galaxies, Springer, New York
\bibitem[]{}
West M.J., 1993, MNRAS 265, 755
\bibitem[]{}
West M.J., C\^ot\'e P., Jones C., et al., 1995, ApJ 453, L77
\bibitem[]{}
Whitmore B.C., Schweizer F., 1995, AJ 109, 960
\bibitem[]{}
Whitmore B.C., 1996, in ``The Extragalactic Distance Scale'' STScI
colloquim proceedings
\bibitem[]{}
Wise R.F.G., Gilmore G., Franx M., 1997, ARA\&A 
\bibitem[]{}
Worthey G., 1994, ApJS 95, 107
\bibitem[]{}
Zepf S.E., Ashman K.M., 1993, MNRAS 264, 611
\bibitem[]{}
Zinn R., 1985, ApJ 293, 424
\bibitem[]{}
Zinn R., 1996, in ``Formation of the Galactic Halo ... Inside and Out'',
ASP Conf.Series, Vol. 92, eds. H.Morrison and A.Sarajedini

\end{thebibliography}
\enddocument